\documentclass[conference]{IEEEtran}
\IEEEoverridecommandlockouts
\usepackage{amsmath,amssymb,amsfonts}
\usepackage{algorithmic}
\usepackage{graphicx}
\usepackage{textcomp}
\usepackage{xcolor}
\def\BibTeX{{\rm B\kern-.05em{\sc i\kern-.025em b}\kern-.08em
    T\kern-.1667em\lower.7ex\hbox{E}\kern-.125emX}}

\usepackage[utf8]{inputenc}
\usepackage{balance}

\usepackage[caption=false, font=footnotesize]{subfig}
\usepackage{booktabs}
\usepackage{xspace}

\usepackage[style=ieee,
            backend=bibtex,                        
            doi=false]{biblatex}
\AtEveryBibitem{\clearfield{pages}}

\newcommand{\figwidth}{1.0\columnwidth}

\begin{document}

\title{Mobile Internet Experience:\\ Urban vs. Rural -- Saturation vs. Starving?}

%\author{}

%\author{
%	\IEEEauthorblockN{Blinded}
%	\IEEEauthorblockA{\textit{Blinded}}
%	Blinded \\
%	Blinded
%}

\author{
	\IEEEauthorblockN{Anika Schwind, Florian Wamser, Tobias Hoßfeld}
	\IEEEauthorblockA{\textit{University of Würzburg, Institute of Computer Science}}
	Würzburg, Germany \\
	\{anika.schwind $|$ florian.wamser $|$ hossfeld\}@informatik.uni-wuerzburg.de
	\and
	\IEEEauthorblockN{Stefan Wunderer, Christian Gassner}
	\IEEEauthorblockA{\textit{Nokia Networks}}
	Ulm, Germany \\
	\{stefan.wunderer $|$ christian.gassner\}@nokia.com
}

\maketitle

\begin{abstract}
%\todo{Paper abstract (between 30 and 300 words)}
Mobile Internet experience has been of increasing interest. 
Services accessed via smartphone applications shall provide satisfying Quality of Experience (QoE), irrespective of end user location, time of the day, and other circumstances. 
Unfortunately, current LTE networks often don't provide constant user throughput, one of the major system influence factors to mobile Internet QoE. 
In this paper, we conducted an exemplary measurement study in LTE networks, comparing the QoE of mobile networks in an urban and a rural region.
Our results show that there are significant differences concerning the network speed which can result in unsatisfactory service quality depending on the application to be used.
When evaluating the QoE for multiple users who are using the same base station in a specific area, user satisfaction decreases drastically, especially in rural areas.
Our work encourages for future work to focus on this gap between the QoE in urban and rural areas.

\end{abstract}


\section{Introduction and Related Work}\label{sec:intro}

Today, the Internet is omnipresent.
People are connected no matter where they are and what they are doing.
At home or work, e.g., stable DSL network connections guarantee a good QoE when using online services.
However, people are often on the move, getting from one place to another by bus, train, or car, crossing rural as well as urban areas, while expecting a similar Internet experience from the mobile networks.
%Here, the QoE is highly influenced by different factors.
The QoE can be influenced by several different factors along the end to end chain.
%On the part of users, according to~\cite{dasari2018impact}, the actual used device is one of them.
%To evaluate this factor, the authors conducted measurements using seven different smartphone types varying in price and hardware.
%While for video applications like video streaming, the performance does not differ much using different smartphonese, the performance of Web browsing is much more sensitive.
Regarding the network, a major QoE influence factor is the network speed.
Here, the minimum, median, and current measured download throughput are highly significant in predicting user satisfaction~\cite{finley2017does}.

As Internet Service Providers (ISP) are interested in measuring their customers' satisfaction with their service, it is necessary to monitor the QoE in their network. 
An overview of state-of-the-art quality monitoring models and measuring methods is given in ~\cite{robitza2017challenges}.
Here, the authors compare different approaches and highlight the major challenges for ISPs in ensuring high service quality.
There already exist studies about the current situation in the mobile networks.
For example, \cite{schwinddissecting} conducted large scale QoE measurements for video streaming in four European countries using a dedicated Docker container within a distributed testbed. %, and they evaluated the impact of parameters from different layers on the streaming quality of experience (QoE).
In contrast, \cite{wu2017mopeye} used a smartphone application to measure per-app mobile
network performance and evaluated the results of their crowdsourcing study on the basis of different factors like %the country distribution or ########## Was ist denn eine Country distribution ? ###########
connection types (WiFi/cellular).
Nevertheless, to the best of our knowledge, no research exists which explicitly faces the type of area in which the QoE measurements are taken, e.g., in a densely populated areas or in a more rural environment.

In this work, a first study was conducted, measuring the mobile network speed in an urban compared to a rural area. 
We evaluated the results with focus on the download throughput and interpreted them according to the resulting QoE of users of this network.
The measurements show that the mobile network provides sufficient QoE for lightweight applications (Google Maps, WhatsApp, Netflix video streaming up to HD) in all measured locations, while in the rural area heavyweight services like 4k video streaming are not possible.
When evaluating the QoE of several competing users, the rural area performs significantly worse than the urban area.
The results show that it is worth to focus on the difference of the network QoE in urban and rural areas, and that further work is required to satisfy all users.

The remainder of this work is structured as follows. 
In Section~\ref{sec:methodology}, the methodology to measure the network speed and the location is described.
Section~\ref{sec:evaluation} deals with the measurement data and evaluates the impact of the maximal available throughput in a given area on the QoE.
Finally, Section~\ref{sec:conclusion} concludes and gives an outlook on future work.

\section{Methodology}\label{sec:methodology}

To compare the current available mobile network quality in urban as well as in rural locations, we conducted a small measurement campaign in two locations in Germany.
Therefore, we ran repetitive network measurements every five minutes.
We used the Docker container MONROE-Nettest~\cite{midoglu2018monroe}, with which we were able to conduct mobile speed measurements and to collect comparable information about the download and upload data rate as well as the median TCP payload round-trip time (RTT). 
The measurements were conducted on a MacBook Air which was connected to the mobile network using USB Tethering and an Android Smartphone equipped with a SIM card with unlimited data plan and no speed caps.

For our measurements, two locations were selected.
The first one, denoted as \textit{urban area}, was in Berlin, Germany.
Here, the measurements were conducted in a  mixed residential and business area in the city center having a typical inter-site distance of 500\,m between base stations. 
The second group of measurements, further mentioned as \textit{rural area}, were performed in a hilly, forested area with very few and small buildings near Cologne.
Here, as the population density is considerably smaller than the density in cities, the distance between base stations is around 5\,km.
Both measurements were performed for more than 24 consecutive hours.
For the urban area, 325 measurements were performed from December 11, 2018, at 8:28\,pm until December 13, 2018, at 8:23\,am.
In the rural area, the network speed was measured 371 times in the period from December 16, 2018, at 2:25\,pm until December 18, 2018, at 5:53\,am.

%The measurements were performed with a mobile app performing up- and downloads regularly via the LTE Internet.
%Within the urban resp. the rural area arbitrary movements with the mobiles were conducted. 
%Both measurements were performed for more than 24 consecutive hours. 
%The urban environment consisted of a  mixed residential and business area in the center of the German capital Berlin. 
%4-5 floor buildings dominate, interrupted by some taller office buildings. 
%The typical inter-site distance is 500m. 
%The rural measurements were performed in a hilly, forested area with very few and small buildings near Cologne. 
%In the examined mobile network, the distance between base stations is around 5km.
 \section{Measurement Results and Discussions}\label{sec:evaluation}

%In this section, the results of our network measurements are presented and compared according to the resulting QoE for different applications.

\subsection{Network measurements: throughput and RTT}

\begin{figure} 
	\centering
	\includegraphics[width=\figwidth]{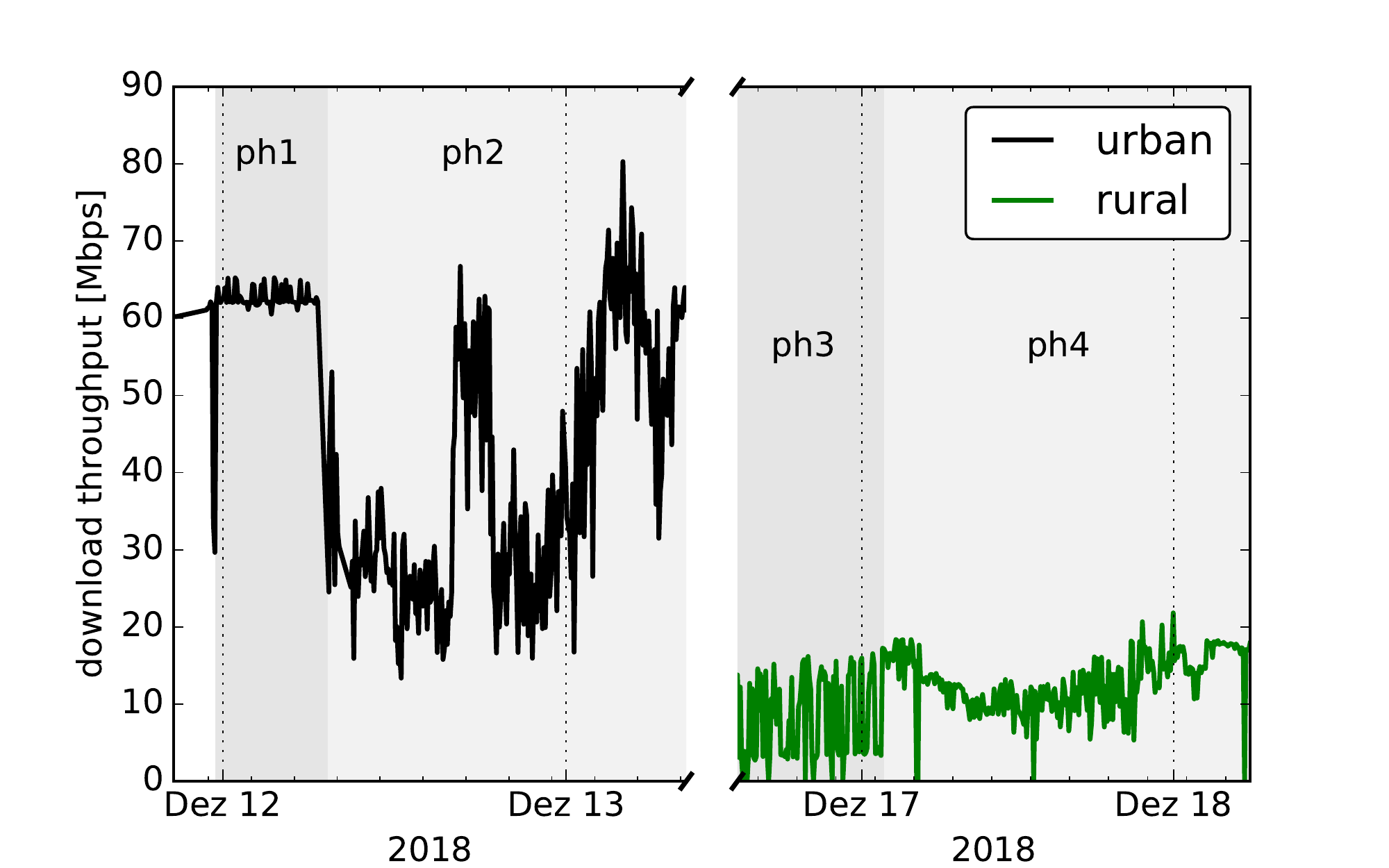}
	\caption{Timeline of the measured download throughput}
	\label{fig:dl_throughput_timeline} 
\end{figure}

First we consider the measured download throughput over time in the two locations in Figure~\ref{fig:dl_throughput_timeline}.
%The x-axis shows the dates while the y-axis indicates the measured throughput in Mbps.
The results marked in black are measured in our urban area, while the results marked in green are measured in our rural area.
The information about the mean, minimum, and maximum value as well as the standard deviation can be found in Table~\ref{tab:stats}.
The urban area values fluctuate between 13.3\, Mbps and 80.31\,Mbps, having a mean download throughput of 45.14\,Mbps.
Here, two phases are visible: phase 1 ($ph1$) between 11:33\,pm and 6:38\,am and phase 2 ($ph2$) between ph2 7:25\,am and 8:23\,am.
In $ph1$, the measured download throughput is relatively stable, having a standard deviation of only 1.15\,Mbps, and, with a mean of 62.67\,Mbps, is very high.
Here, the measurements were conducted at a fixed location at home, having good signal strength.
The traffic during the night was probably low, resulting in a high constant throughput.
In $ph2$, during a long business day, measuring in various urban locations with wide variations in signal strength and traffic result in a higher throughput fluctuation, having a standard deviation of 16.18\,Mbps.
Nevertheless, the mean throughput is still high (26.90\,Mbps).
In contrast to that, the measured throughput in the rural area is significantly smaller, varying between no throughput at all and a maximum of 21.81\,Mbps, having a mean download throughput of 11.44\,Mbps.
In the rural area, no clear variation depending on the daytime is visible, probably due to low traffic variations and due to low signal strength as dominating limiting factor.
Thus, we split the results into two phases according to their fluctuations: phase 3 ($ph3$) between 2:25\,pm and 1:41\,am and phase 4 ($ph4$) between ph2 1:48\,am and 5:53\,pm.
The standard variation of 5.41\,Mbps in $ph3$ is lower than in $ph2$, but also the mean throughput is much lower.
As $ph4$ is relatively stable, showing only a few peaks, the standard deviation is only 3.78\,Mbps.

Not only concerning the download throughput, but also in terms of the upload throughput and the RTT, the results in the urban area are significantly higher than in the rural area.
Table~\ref{tab:stats} shows the statistics of upload throughput and RTT.
The upload throughput (Up T), for example, shows dramatic differences, having a mean of 23.85\,Mbps in the urban and only 0.22\,Mbps in the rural area.

\begin{table}
	\centering
	\caption{Statistics of the collected network measurements}
	\label{tab:stats}
	\begin{tabular}{lrrrr}
		\toprule
		                       &  Mean &   Min &      Max &   SD \\ \midrule
		Dl T urban (Mbps)      & 45.14 & 13.36 &    80.31 & 17.19 \\
		Dl T urban ph1 (Mbps)  & 62.67 & 60.53 &    65.25 &  1.15 \\
		Dl T urban ph2 (Mbps)  & 39.45 & 13.36 &    80.31 & 16.18 \\ \midrule
		Dl T rural (Mbps)      & 11.44 &  0.00 &    21.81 &  4.88 \\
		Dl T rural  ph3 (Mbps) &  7.93 &  0.00 &    17.24 &  5.41 \\
		Dl T rural ph4 (Mbps)  & 12.87 &  0.00 &    21.81 &  3.78 \\ \midrule
		Up T urban (Mbps)      & 23.85 &  0.00 &    39.03 &  5.48 \\
		Up T rural (Mbps)      &  0.22 &  0.00 &     1.42 &  0.32 \\ \midrule
		RTT urban (ms)         & 63.35 & 48.53 &   724.55 & 43.57 \\
		RTT rural (ms)         & 89.26 & 65.73 & 1,833.84 & 92.98 \\ \bottomrule
	\end{tabular}
\end{table}

\begin{figure} 
	\centering
	\includegraphics[width=\figwidth]{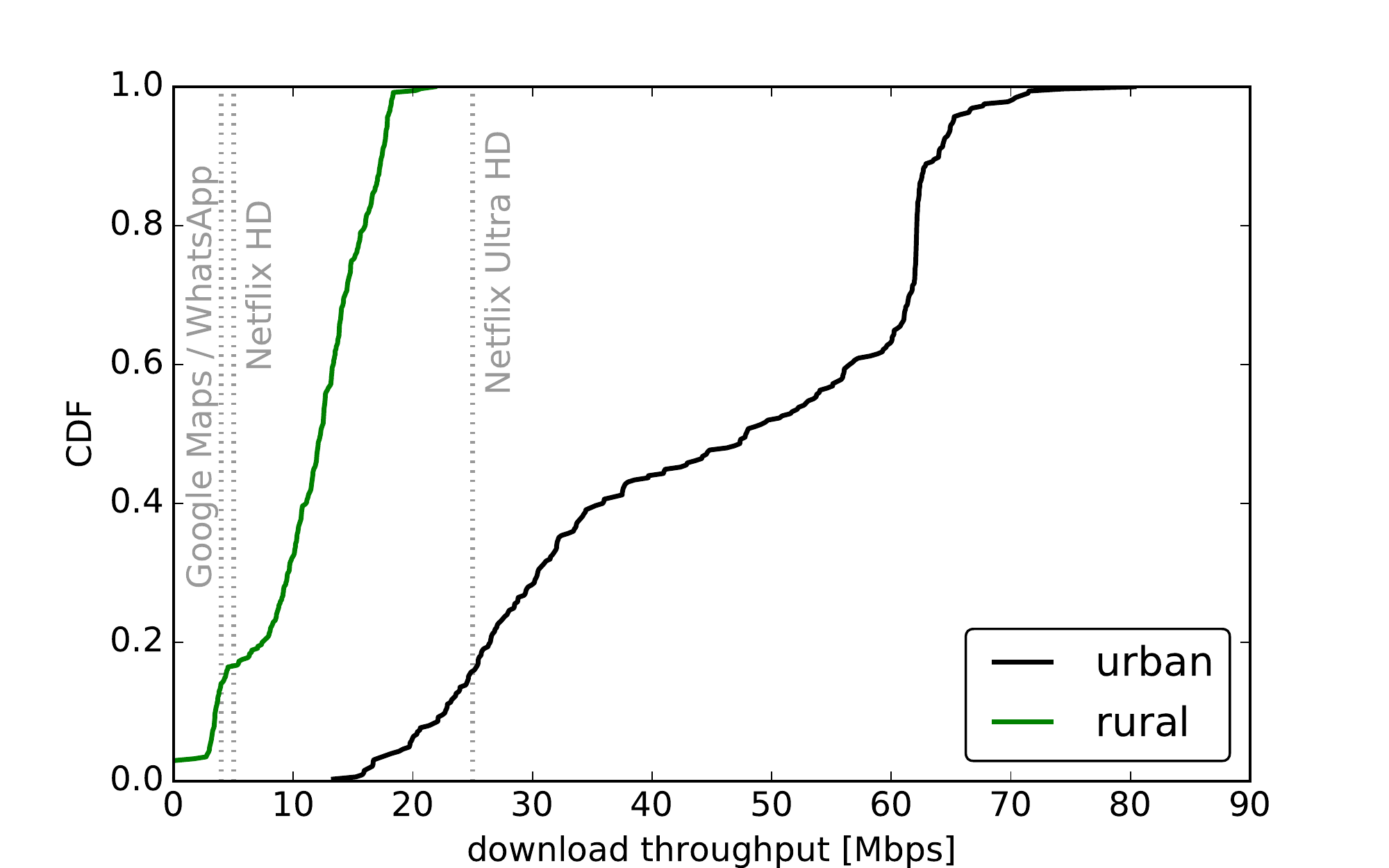}
	\caption{Complementary distribution function of the download throughput}
	\label{fig:dl_throughput_cdf} 
\end{figure}

\subsection{User experience: Which applications can be served?}
The question arises which apps can be served with high QoE in the urban and rural areas. To this end, the measured network throughput is compared to the throughput requirements for different applications yielding a good QoE.
Figure~\ref{fig:dl_throughput_cdf} shows the empirical complementary cumulative distribution function (CDF) of the measured throughput in the urban and rural areas. 
Throughput thresholds resulting in good QoE for popular smartphone applications are drawn as gray dashed lines. The required throughput is
4\,Mbps for Google Maps and the messaging application WhatsApp~\cite{casas2015exploring}.
%The first threshold is set at 4\,Mbps, indicating the required download throughput for a good QoE for Google Maps and the messaging application  WhatsApp~\cite{casas2015exploring}.
%Next, different thresholds for video streaming are shown.
Netflix provides the following recommendations for video streaming\footnote{Internet connection speed recommendation for different video resolutions: \url{https://help.netflix.com/en/node/306} (Accessed: 2018-03-08)} to avoid stalling: % instead of \cite{netflixbandwidth}
3\,Mbps for SD; 5\,Mbps for HD, 25\,Mbps for Ultra HD videos.
As the throughput in the rural area was very low, with a probability of 14.29\,\%, the bandwidth would be not sufficient to use Google Maps or WhatsApp and providing a good user experience.
Having a look at Netflix video streaming, in 16.71\,\% of the cases the available bandwidth would be too low to reach a high user satisfaction for streaming an HD video.
Thinking about 4K videos, which are, for example, used for VR video streaming, it would be impossible to stream them with high QoE.
In an urban area, smaller apps like Google Maps and WhatsApp can be used without any problem.
The same holds for HD video streaming.
Contrary to the rural area, in the urban area in most of the cases (84.00\,\%) even streaming a Ultra HD video would be possible without stalling.

\subsection{User experience: Competing users in mobile networks}

\begin{figure} 
	\centering
	\includegraphics[width=0.95\columnwidth]{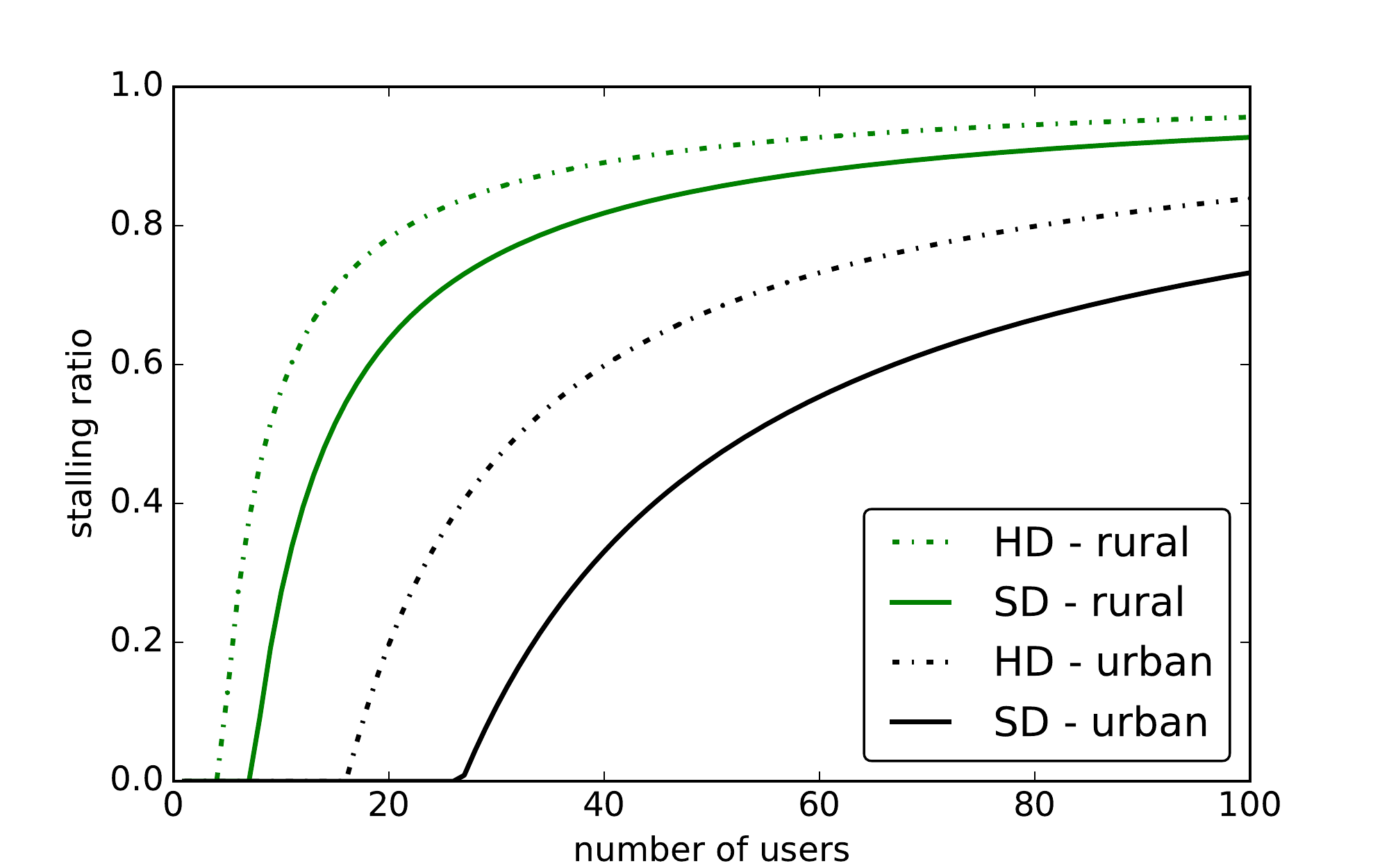}
	\caption{Stalling ratio for number of users who simultaneously stream videos}
	\label{fig:user_sim_video} 
\end{figure}

Based on the network measurement results, we conducted now a simulation study to investigate the user perceived quality when there are several competing users in the network. To this end, we considered video streaming being one of the most popular Internet applications in the past decades in the mobile Internet. % and people increasingly stream videos using the mobile network, it becomes important for ISPs to measure their customers QoE with this service.
%As video streaming has become one of the most popular Internet applications in the past decades and people increasingly stream videos using the mobile network, it becomes important for ISPs to measure their customers QoE with this service.
According to different subjective studies, e.g., \cite{hossfeld2011quantification,mok2011measuring}, stalling is one of the KPIs for video streaming. For mobile operators and service providers, it is important to know whether the bandwidth is high enough to prevent stalling.
%According to~\cite{hossfeld2015each}, 
As metric we use the stalling ratio $R$ which quantifies the probability that a video stalls. The stalling ratio is derived analytically in \cite{hossfeld2015each} and depends on the ratio between available network bandwidth $B$ and video bitrate $V$, $R=1-B/V$.
%\begin{eqnarray}
%R &=& 1 - \frac{available \; bandwidth}{video \; bitrate}
%\end{eqnarray}
In Figure~\ref{fig:user_sim_video}, we calculated the stalling ratio for the maximum measured bandwidth in the urban 80.31\,Mbps (marked in black) and rural area 21.81\,Mbps (marked in green) for different numbers of users simultaneously watching videos of the same quality.
We assume that the available bandwidth is evenly distributed over all users on average.
The x-axis shows the number of users streaming videos at the same time in the same cell, while the y-axis indicates the stalling ratio.
A clear trend is visible: For the urban use cases, the stalling ratio is considerably smaller than for the rural use cases.
For example, having a look at SD videos, in the urban area, up to 26 users can stream a video without stalling, while only 7 users are served without video interruptions in a rural area.
By increasing the video quality to HD videos, in urban areas 16 users can stream the video without stalling, while in rural areas the stalling ratio for 16 users would be 72.73\,\% and thus, unacceptable.

%%urban
%SD - 26
%HD - 16
%
%%rural
%SD - 7
%HD - 4

\section{Conclusion and Future Work}\label{sec:conclusion}

In this paper, we conducted network measurements in an urban as well as in a rural area to answer the question whether the mobile network QoE in both areas is satisfactory.
We found that especially the download throughput in the rural area was significantly lower than in the urban region.
This results in the fact that for one single user, the service quality is sufficient only for lightweight applications in our rural mobile network.
%However, thinking about multiple users who stream, for example, videos at the same time, the user satisfaction decreases drastically.
However, the network usage at busy hour is one factor which can lead to reduced throughput in the whole end to end chain and thus, can lead to low user satisfaction.
This study should give an incentive for future work to focus this gap between the QoE in urban and rural areas.
Therefore, our measurements should be repeated in a larger scale to increase the validity and provide more detailed conclusions.

%\section*{Acknowledgment}
% \emph{- Left out due to double-blind submission. -}
%  This work was partly funded in the framework of the EU ICT project MONROE (H2020-2014-ICT-644399, through open call project Mobi-QoE). The authors alone are responsible for the content.

\balance
\renewcommand*{\bibfont}{\small}
\printbibliography

@inproceedings{hossfeld2011quantification,
	title={Quantification of YouTube QoE via crowdsourcing},
	author={Ho{\ss}feld, Tobias and Seufert, Michael and Hirth, Matthias and Zinner, Thomas and Tran-Gia, Phuoc and Schatz, Raimund},
	booktitle={IEEE International Symposium on Multimedia (ISM)},
	pages={494--499},
	year={2011},
	_organization={IEEE}
}

@inproceedings{mok2011measuring,
	title={Measuring the quality of experience of HTTP video streaming.},
	author={Mok, Ricky KP and Chan, Edmond WW and Chang, Rocky KC},
	booktitle={Integrated Network Management},
	pages={485--492},
	year={2011}
}

@inproceedings{casas2015exploring,
	title={{Exploring QoE in cellular networks: How much bandwidth do you need for popular smartphone apps?}},
	author={Casas, Pedro and Schatz, Raimund and Wamser, Florian and Seufert, Michael and Irmer, Ralf},
	booktitle={5th Workshop on All Things Cellular: Operations, Applications and Challenges},
	pages={13--18},
	year={2015},
	organization={ACM}
}

@inproceedings{hossfeld2015each,
	title={To each according to his needs: Dimensioning video buffer for specific user profiles and behavior},
	author={Ho{\ss}feld, Tobias and Moldovan, Christian and Schwartz, Christian},
	booktitle={2015 IFIP/IEEE International Symposium on Integrated Network Management (IM)},
	pages={1249--1254},
	year={2015},
	_organization={IEEE}
}

@article{finley2017does,
	title={Does network quality matter? {A} field study of mobile user satisfaction},
	author={Finley, Benjamin and Boz, Eren and Kilkki, Kalevi and Manner, Jukka and Oulasvirta, Antti and H{\"a}mm{\"a}inen, Heikki},
	journal={Pervasive and Mobile Computing},
	volume={39},
	pages={80--99},
	year={2017},
	publisher={Elsevier}
}

@inproceedings{wu2017mopeye,
	title={Mopeye: Opportunistic monitoring of per-app mobile network performance},
	booktitle={2017 USENIX Annual Technical Conference (USENIX ATC'17)},
	author={Wu, Daoyuan and Chang, Rocky KC and Li, Weichao and Cheng, Eric KT and Gao, Debin},
	pages={445--457},
	year={2017}
}

@article{robitza2017challenges,
	title={Challenges of future multimedia {QoE} monitoring for internet service providers},
	author={Robitza, Werner and Ahmad, Arslan and Kara, Peter A and Atzori, Luigi and Martini, Maria G and Raake, Alexander and Sun, Lingfen},
	journal={Multimedia Tools and Applications},
	volume={76},
	number={21},
	pages={22243--22266},
	year={2017},
	publisher={Springer}
}

@inproceedings{midoglu2018monroe,
	title={{MONROE-Nettest}: A configurable tool for dissecting speed measurements in mobile broadband networks},
	author={Midoglu, Cise and Wimmer, Leonhard and Lutu, Andra and Alay, {\"O}zg{\"u} and Griwodz, Carsten},
	booktitle={IEEE INFOCOM 2018-IEEE Conference on Computer Communications Workshops (INFOCOM WKSHPS)},
	pages={342--347},
	year={2018},
	_organization={IEEE}
}

@article{schwinddissecting,
	title={Dissecting the performance of YouTube video streaming in mobile networks},
	author={Schwind, Anika and Midoglu, Cise and Alay, {\"O}zg{\"u} and Griwodz, Carsten and Wamser, Florian},
	journal={International Journal of Network Management},
	volume={e2058},
	year = {2019},
	publisher={Wiley Online Library}
}
%\bibliographystyle{IEEEtran}
%\bibliography{sections/references} 

\end{document}